\documentstyle[12pt,aaspp4,flushrt]{article}

\newcommand{\bmv}{({\bv})}
\newcommand{\civ}{{\rm C}\thinspace{\sc{iv}}}
\newcommand{\eg}{{\it e.g.}}
\newcommand{\etal}{{\it et al.\/}}
\newcommand{\feii}{Fe\thinspace{\sc{ii}}}

\newcommand{\mgii}{Mg\thinspace{\sc{ii}}}
\newcommand{\msun}{$M_{\odot}$}
\newcommand{\oii}{[O\thinspace{\sc{ii}}]}
\newcommand{\oiii}{[O\thinspace{\sc{iii}}]}
\newcommand{\vmi}{(\hbox{$V\!-\!I$})}
\newcommand{\vmn}{(\hbox{$V\!-\!N$})}

\begin{document}

% Modified aaspp4.sty 2 June 1997 so that bbox is def'd where boldmath is used
%   in the present title

\title{High-{$\boldmath z$} Ly{$\boldmath\alpha$} Emitters. I. A Blank-Field 
  Search for Objects Near Redshift {$\boldmath z$} = 3.4 in and around the 
  Hubble Deep Field and the Hawaii Deep Field SSA22\altaffilmark{1}}
\author{Lennox L. Cowie\altaffilmark{2,3} and Esther M. Hu\altaffilmark{2}}
\affil{Institute for Astronomy, University of Hawaii, 2680 Woodlawn
  Drive, Honolulu, HI 96822\\
  cowie@ifa.hawaii.edu, hu@ifa.hawaii.edu}
\altaffiltext{1}{Based in part on observations with the NASA/ESA {\it Hubble 
  Space Telescope} obtained at the Space Telescope Science Institute, which 
  is operated by AURA, Inc., under NASA contract NAS 5-26555.}
\altaffiltext{2}{Visiting Astronomer, W. M. Keck Observatory,
  jointly operated by the California Institute of Technology, the University
  of California, and the National Aeronautics and Space Administration.}
\altaffiltext{3}{Visiting Astronomer, Canada-France-Hawaii Telescope,
  operated by the National Research Council of Canada, the Centre National
  de la Recherche Scientifique of France, and the University of Hawaii.}

\begin{abstract}
We present deep narrow-band ($\lambda = 5390$ \AA, $\Delta\lambda
= 77$ \AA) and multi-color observations of the Hubble Deep Field
and the Hawaii Deep Field SSA22 obtained with the LRIS instrument
at the Keck\thinspace{II} 10-m telescope.  It is shown that there
is a substantial population of galaxies at $z\sim3.4$ which can be
selected by Ly$\alpha$ emission.  Comparison with color-selected samples 
shows that the samples selected with these different criteria have
substantial, but not complete overlap, and that there is a
comparable surface density in the two selected populations.  The 
emission-line selected samples include objects with strong Ly$\alpha$,
and which are significant contributers to the integrated star
formation at these epochs. For a Salpeter IMF we estimate a minimum
star formation rate of 0.01 \msun\ Mpc$^{-3}$ yr$^{-1}$ at $z=3.4$
for $H_0$ = 65 km s$^{-1}$ Mpc$^{-1}$ and $q_0$ = 0.5 in the
Ly$\alpha$-selected objects, though the value could be substantially higher
if there is significant extinction.
\end{abstract}


\section{Introduction}

Recent opinion has swung heavily to the viewpoint that
emission-line searches for high-$z$ Ly$\alpha$ emitters have failed,
and that color-based searches may represent the only realistic
way to identify high-redshift galaxies.  In the  present series
of papers we shall show that this assumption is ill-founded and
that slightly more sensitive searches than previously carried out
yield very large numbers of high-$z$ Ly$\alpha$ emitters, including
some objects which are so faint in the continuum that they would
not be easily detected on the basis of their colors.  The results
show that emission-line searches using 10-m class telescopes and
either narrow-band or spectroscopic techniques are an efficient
means of identifying and studying the distribution and clustering
of high-redshift galaxies, including objects beyond redshift 5.

Indeed, the pessimism has not been justified even based on existing
data.  Surveys around known objects have yielded numbers of
Ly$\alpha$ emitters both at moderate $z$ (Djorgovski
\etal\markcite{djorg85} 1985; Macchetto \etal\markcite{macch} 1993;
Francis \etal\markcite{francis} 1996; Pascarelle
\etal\markcite{pasc96a} 1996; Djorgovski \etal\markcite{djorg_dla}
1996; Francis \etal\markcite{francis97} 1997) and high $z$ (Hu
\etal\markcite{br1202} 1996; Petitjean \etal\markcite{petit96} 1996;
Hu \& McMahon\markcite{br2237} 1996; Hu \etal\markcite{hx37_hu}
1997), and while these results may be dismissed as a consequence of
the possibly anomalous environments around the target objects, it is
also true that quite a large fraction of the color-selected
$z=2\to4$ objects have Ly$\alpha$ in emission (cf. Steidel
\etal\markcite{stei96a} 1996a, \markcite{stei96b}1996b; Lowenthal
\etal\markcite{low97} 1997; Steidel \etal\markcite{stei98} 1998).
Indeed, these latter results show at once that blank field objects
can be found with only a small increase in sensitivity over the
existing Ly$\alpha$ surveys summarized by Pritchet\markcite{pri94} (1994).

A substantial gain in depth is now possible with the advent of the
10-m telescopes, and in order to exploit this we have begun a
narrow-band filter search for high-$z$ Ly$\alpha$ emitters using the LRIS
instrument (Oke \etal\markcite{lris} 1995) in imaging mode on the
Keck\thinspace{II} 10-m telescope with specially designed filters.
In the present paper we describe observations of a 25.3 $\sq'$
region surrounding the Hubble Deep Field (HDF)(Williams
\etal\markcite{hdf} 1996) and a similar area around the Hawaii Deep
Field SSA22 (Cowie \etal\markcite{cowie_1} 1994) in broad-band $B$,
$V$, and $I$ colors, and using a filter centered at $5390\,$\AA\ which
covers a $77\,$\AA\ bandpass.  This filter would detect Ly$\alpha$ at
$z=3.405 \to 3.470$.  Subsequent papers in this series will address
deep slit spectroscopy of quasar and blank fields, and deep optical
and infrared narrow-band imaging studies at higher redshifts.  At the
time of the filter design no objects were known to lie within the
filter redshift range, but subsequently one such object was
identified by Lowenthal \etal\markcite{low97} (1997) in the HDF field
itself.  The search is therefore a blank-field search, but among
the objects identified as Ly$\alpha$ emitters in this redshift interval,
the survey does indeed recover the $z=3.430$ object found by Lowenthal 
\etal\markcite{low97} (identified as hd2-0698-1297 in their paper).

SSA22 and the HDF are ideal test areas for a narrow-band imaging
search because of the large number of spectroscopic redshift
identifications in and around these areas and the ultra-deep
multicolor photometric information available in the HDF proper.
Thus a survey using these fields allows us to address issues such
as: (1) comparing methodologies and sensitivities of narrow-band
emission-line techniques and continuum color-break techniques for
identifying high-redshift galaxies, (2) determining relative
numbers of foreground emission-line objects and determining the
best methods for distinguishing between the various types of
emitters (\eg, Ly$\alpha$, \oii, H$\beta$, \oiii), (3) estimating the
relative surface density of high-redshift candidates as a
function of flux found using each technique, and (4) establishing
a baseline for comparison between blank-field and targeted
searches around high-$z$ objects (such as radio galaxies, DLAs,
and quasars), and for evolution of galaxy properties at higher
redshifts.  In addition, we can establish specified criteria
(\eg, desired magnitude limits, required precision of color
estimates for separating objects of different redshift, typical
spatial extents) on well-studied fields with high-resolution
images, which then permits fine-tuning future investigations of
high-redshift galaxies.  Finally, when used in conjunction with
optical and infrared narrow-band surveys currently in progress,
the incidence and statistics of foreground emission-line objects
are also of interest for estimating the surface densities of
star-forming galaxies and AGN at different epochs, and for
comparing indices such as \oii\ with primary star-formation
indices such as H$\alpha$ (Kennicutt\markcite{kenn83}\markcite{kenn92}
1983, 1992) or the far UV flux, in order to track the evolution of 
star formation history to very high redshifts. We shall address 
these issues in subsequent papers.

\section{Data}

Narrow-band observations using the LRIS camera were obtained
on the Hubble Deep Field (HDF) on the night of UT 1997 May 2 and on 
SSA22 on the nights of UT 1997 August 8 and 10.  The narrow-band
filter was a specially designed $9\farcs5$ square interference
filter centered at $5390\,$\AA, with a peak transmission of just
under 70\% and FWHM of $77\,$\AA, and located in the collimated beam
at the standard LRIS filter position.  The filter lies in a very
dark region of sky and was also matched to a dark portion of the
IR sky where \oii\ would lie if Ly$\alpha$ were in the optical
narrow-band.  For the HDF a series of 900-sec exposures (2 hrs)
shifted by 15$''$ between successive frames were taken in the
narrow-band filter, immediately followed by a series of 360-sec
$V$-band exposures (0.6 hrs) on the same field.  These data were
combined with a series of 420-sec $B$-band exposures (0.7 hrs)
taken with LRIS on the night of UT 1997 March 8.  For SSA22 a 5
hr narrow-band exposure was taken, again as a series of 900-sec
exposures, while the $B$ data were acquired as a series of
420-sec exposures (0.8 hrs) on UT 1997 August 9, with the $V$
data taken as a series of 400-sec exposures over the nights of UT
1997 August 8 and 10 (1.6 hrs), and $I$ data taken as a series of
300-sec exposures on UT 1997 August 10 (0.7 hrs).  The data were
processed using median sky flats generated from the dithered
images and calibrated using observations of Landolt standard
stars (Landolt\markcite{landolt} 1992) and spectrophotometric
standards (Massey \etal\markcite{massey} 1988). The FWHM is
$\sim0\farcs7$ on all the composite images, which were obtained
under photometric conditions, and for the HDF the narrow-band
reaches a $1\ \sigma$ limiting $AB$ magnitude of 26.8 in a $3''$
diameter aperture, corrected to a total magnitude following the
procedures of Cowie \etal\markcite{cowie_1} (1994).  The
corresponding $1\ \sigma$ flux limit is $6 \times 10^{-18}\ {\rm
erg\ cm}^{-2}\ {\rm s}^{-1}$ over a $3''$ diameter aperture.  For
SSA22 the $1\ \sigma$ limit is 0.7 mags fainter, and the flux
limit is $3 \times 10^{-18}\ {\rm erg\ cm}^{-2}\ {\rm s}^{-1}$.

For the HDF additional $V$ and $I$ data were obtained with the
UH8K camera at CFHT (Barger \etal\markcite{barger} 1998) on the
nights of UT 1997 April 3--8.  These comprise twenty-two 1200-sec
$I$-band exposures  (8.3 hrs) and thirty-three 1200-sec $V$-band
exposures (11 hrs) under conditions of mixed transparency.  Net
image quality is $\sim0\farcs8$ FWHM in $I$ and $\sim0\farcs9$ in $V$
on these exposures. The UH8K $V$ data were registered to the
LRIS data and a noise-weighted addition was performed.  The
$I$-band data were calibrated using shallower images obtained
under photometric conditions with the UH 2.2 m telescope.  The
corresponding $1\ \sigma$ limits for the combined HDF observations
over a corrected $3''$ diameter aperture are 27.2 ($B$), 27.4 ($V$), 
and 25.8 ($I$), where $B$ and $V$ are on the Johnson system and $I$ 
is Kron-Cousins. For SSA22 the corresponding limits are 27.8 ($B$),
27.5 ($V$), and 25.8 ($I$).

Two catalogs were next generated.  The first was a narrow-band catalog
of all objects selected to lie above a surface brightness of 27.9
mags/$\sq''$ ($AB$) in a $0\farcs8$ boxcar-smoothed narrow-band image,
and above a $5\ \sigma$ limiting aperture magnitude $N$($AB$) = 25 for 
the HDF and 25.7 for SSA22.  $B$, $V$, and $I$ magnitudes were
measured for all these objects, and redshifts compiled from the
literature (Songaila \markcite{ksurvey_3} 1994; Steidel
\etal\markcite{stei96a} 1996a; Cowie \etal\markcite{large_sample} 1996;
Cohen \etal\markcite{cohenhdf} 1996; Lowenthal \etal\markcite{low97}
1997; Steidel \etal\markcite{stei98} 1998) and from unpublished spectroscopy
by our group, which may be found in the Hawaii Active Catalog of the HDF 
(Songaila \etal\markcite{hdf_active} 1997).  The combined narrow-band
catalogs contain 1574 objects brighter than $N(AB) = 25$ of which 412
have spectroscopic identifications, and for SSA22 an additional 286 objects 
with $25 < N(AB) < 25.5$ of which only ten have spectroscopic identifications.
A second $V$-selected catalog was generated consisting of all objects with
surface brightnesses above 29.1 mags/$\sq''$ in the $0\farcs8$ smoothed
$V$ image and with $V < 25.7$ (the $5\ \sigma$ limit for the HDF).
$B$, $V$, and $N(AB)$ magnitudes were measured for these objects, and
redshifts were compiled.  The $V$-selected catalog contains 2375
objects, of which 457 have spectroscopic IDs.  Because the center of
the narrow-band filter lies very close to the center of the $V\/$
filter there is only a very small color difference term.  In the
subsequent discussion we use $N = N(AB) + 0.11 - 0.04$\vmi\ to correct
for this.  The magnitudes in the SSA22 field have also been corrected
for a small amount of galactic extinction ($E_{B-V} = 0.05$) using a
standard reddening law.

\begin{figure}[h]
\figurenum{2b}
\epsscale{.75}
\plotone{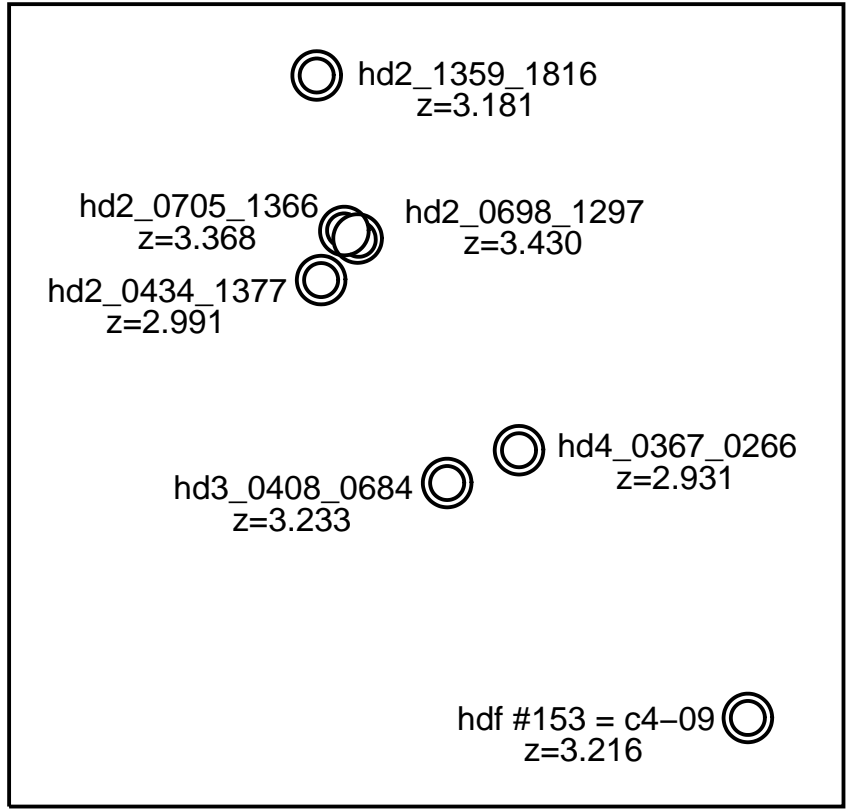}
\caption{
Identifications of high-redshift galaxies in the HDF which have been
spectroscopically identified in Lowenthal \etal\protect{\markcite{low97}}
1997, in the Hawaii Active Catalog\protect{\markcite{hdf_active}}, or in
Steidel \etal\protect{\markcite{stei96a}} 1996a.\label{fig:2b}
}
\end{figure}

In Figs.~\ref{fig:1a} and \ref{fig:1b} we show the LRIS fields of the 
narrow-band exposures -- over a $380'' \times 275''$ region for the HDF 
(Fig.~\ref{fig:1a}) and over a $390'' \times 280''$ region for SSA22 
(Fig.~\ref{fig:1b}).  The catalog field corresponds to objects within the 
central $6\farcm0 \times 4\farcm2$ region, providing similar coverage to 
other $z > 4$ Ly$\alpha$ narrow-band surveys around quasars (\eg, Hu \& 
McMahon\markcite{br2237} 1996) that will be discussed in later papers, or an 
area roughly $4.4 \times$ the size of the HDF proper.  Objects with observed 
equivalent widths in excess of 77 \AA\ in the narrow-band catalog are circled;
objects which additionally have measured redshifts within this sample are
circled with heavy borders (here, just the $z=3.430$ object in the HDF
found by Lowenthal \etal\markcite{low97} 1997 and shown in Fig.~\ref{fig:1a}).  
Image ghosts can be seen around the brightest objects in the fields but 
do not significantly change the sampling area or sensitivity limits.  
Figure~\ref{fig:2a} shows the corresponding regions for the Keck $V$-band 
images of the HDF (redshift identifications given in the 
annotations to the overlay in Fig.~\ref{fig:2b}) and SSA22 (Fig.~\ref{fig:2c}). 
In these images circles indicate the $V < 25.5$ objects which are red in \bmv\ 
and blue in \vmi\ (\thinspace\bmv\ $> 1.1$, \vmi\ $ < 1.6$).  Again, objects 
encircled with heavy borders in each figure have measured redshifts, in 
addition to matching the specified color criteria.  In the two fields eight of 
these objects are $z>2.9$ galaxies (including one quasar), four are galaxies
near $z\sim0.3$ and 2 are stars.  The HDF is richer in such objects than
SSA22 by a factor of roughly 2.5, suggesting that there is substantial
variation in the surface density of $z>3$ objects from field to field for
images with areas of this size.

\section{Emission-Line Objects Selected by the Narrow-Band Filter}

\begin{figure}[h]
\stepcounter{figure}
\stepcounter{figure}
\epsscale{0.90}
\plotone{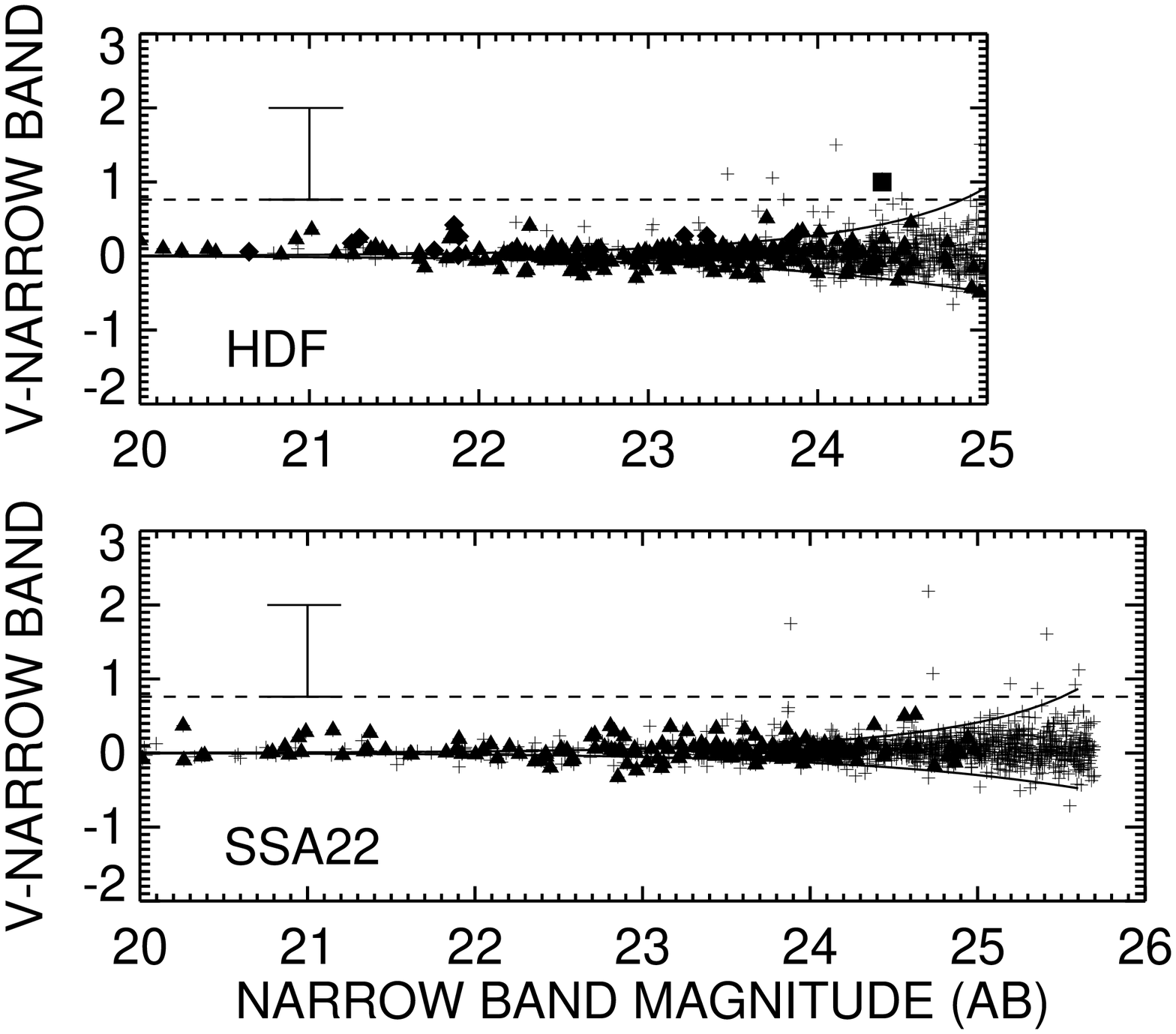}
\caption{The distribution of emission-line excess
\vmn\ vs.\ $N\/$ for objects (pluses) in the narrow-band catalogues for
the HDF and SSA22.  The solid lines show the distribution of $3\ \sigma$
errors. The solid box shows the Lowenthal \etal\protect{\markcite{low97}}
$z=3.430$ object, the diamonds objects where \oii\ 3727 falls in the filter 
bandpass, and the triangles other objects with measured redshifts.  The dashed 
line shows the \vmn\ color corresponding to an equivalent width of 77 \AA\, 
and the vertical bar shows the expected \vmn\ distribution for objects with
observed equivalent widths ranging from 77 \AA\ to $\infty$.\label{fig:3}}
\end{figure}

In Fig.~\ref{fig:3} we plot \vmn\ vs $N$ for the objects in the narrow-band
selected catalog.  The solid lines show the $3\ \sigma$ errors.  The dashed
line shows the \vmn\ color for objects which would have Ly$\alpha$ equivalent
widths in excess of $77\,$\AA\ (that is, in which the flux in the narrow
band is double that of the continuum).  The vertical bar shows the expected
\vmn\ range for objects with observed equivalent widths from $77\,$\AA\ to
$\infty$.  The large solid box shows the Lowenthal \etal\ object.  In the
HDF there are five objects in the field with observed equivalent widths in
excess of $77\,$\AA\ and with $N < 25$, and in SSA22 there are seven such 
objects with $N \leq 25.5$.  The fluxes, equivalent widths, $V$ magnitudes, and
colors of the objects are summarized in Table \ref{tbl-1}.  The equivalent
widths of most of the objects are such that they are unlikely to be
\oii\ emitters since the rest frame \oii\ equivalent width rarely exceeds
$100\,$\AA\ (\eg, Songaila \etal\markcite{ksurvey_3} 1994), and indeed, for
the one spectroscopically identified object the emission is due to Ly$\alpha$,
which is the only plausible alternate.  We therefore consider these objects
as the candidate sample of strong Ly$\alpha$ emitters and attempt to confirm 
this with other diagnostics.

\begin{figure}
\vskip-0.5in
\vbox to1.5in{\rule{0pt}{1.5in}}
\includegraphics{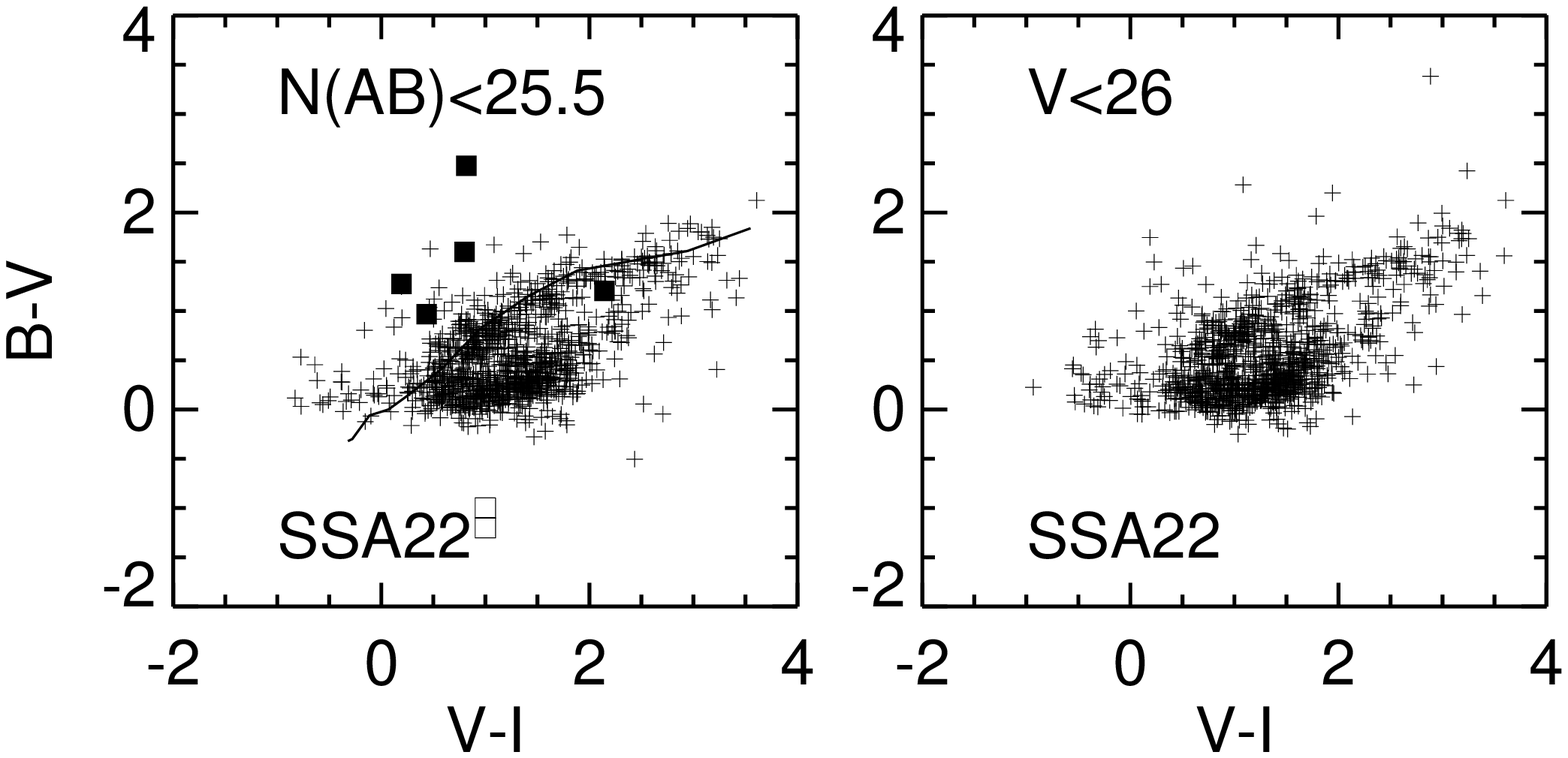}
%\refstepcounter{figure}               % turn off Fig. label
\vskip -0.2in%\label{fig:anonymous}
\caption{\bmv\ vs \vmi\ color-color plots for objects in the
HDF and SSA22 fields selected from the narrow-band and $V-$band
catalogues.  The left-hand plots show the color distributions of objects
in the narrow-band catalogues ($N(AB) < 25$ for the LRIS HDF field and
$N(AB) < 25.5$ for the LRIS SSA22 field).  The colors of emission-line
objects in these magnitude-selected catalogues with equivalent widths in
excess of 77 \AA\ in the observed frame are indicated with squares, and
may be seen to lie at red \bmv\ and blue \vmi\ colors.  Two of the strong
emission-line objects in the SSA22 field have continuua which are either
undetected or too faint to provide color measurements (Table 1).  They are
indicated schematically by open squares placed at nominal color
positions.  The slightly higher density of stars in the SSA22 field
($b^{II}\sim-44^{\circ}$) compared with the HDF ($b^{II}\sim+55^{\circ}$)
may be seen in the more populated star track ({\it solid line}) which
approaches the blue end of the high-$z$ galaxy color distribution.  The
right-hand plots show the color-distribution of objects from our
$V-$selected catalogues ($V<25.5$ for the HDF and $V<26$ for SSA22).  For
the HDF, the available $HST$ filter measurements provide a tighter color
distribution, and here we have elected to use (F602W--F814W) in place of
\vmi\ and (F450W--F602W) in place of \bmv, with the $V < 25.5$ sample
restricted to the HDF proper (about a quarter of the size of the larger
LRIS field).  These points are shown as open triangles.  The steeper slope
of this color distribution is a reflection of the different wavelength
centers (\eg, F602W vs $V$) used for the abscissa.\label{fig:4}}
\end{figure}

In Fig.~\ref{fig:4} we plot the locus of the \vmi\ vs \bmv\ colors for
the narrow-band selected objects in the LRIS fields, and compare these
to the color distribution of $V$-selected catalog objects in the
HDF proper, and $V$-selected objects in the LRIS image of the SSA22
field.  The objects that meet our equivalent width criterion and which
are bright enough in the continuum for colors to be measured (Table \ref{tbl-1})
({\it solid squares}) have predominantly red  \bmv\ colors, but are
blue (or close to flat-spectrum $f_{\nu}$) in \vmi\ colors, as would be
expected for high-redshift objects that are dominated by star formation
(Cowie\markcite{cowie88} 1988; Songaila, Cowie, \& Lilly\markcite{scl}
1990; Steidel \etal\markcite{stei96a}\markcite{stei96b} 1996a, 1996b;
Lowenthal \etal\markcite{low97} 1997).  These objects occupy the same
region of the color-color diagram as objects seen in the tail of the
color-color distribution for the HDF proper at bluer \vmi\ (or F602W -- F814W)
and red \bmv\ (Fig.~\ref{fig:4}, {\it top right panel}), and which correspond 
to high-redshift galaxies selected by color.  As we shall demonstrate
using the colors of HDF objects with known redshifts, these continuum
colors can also distinguish redshift $z\sim3.4$ objects from lower-$z$
emitters.  (We also note that the narrow-band catalog has a shallower
slope in the Fig.~\ref{fig:4} color-color diagrams for the HDF because the 
F602W filter is redder than the $V$-band used with LRIS.)\ \ The color selection
criterion is therefore consistent with the interpretation that nearly
all of the sample are Ly$\alpha$ emitters.

\begin{figure}[h]
\epsscale{0.90}
\plotone{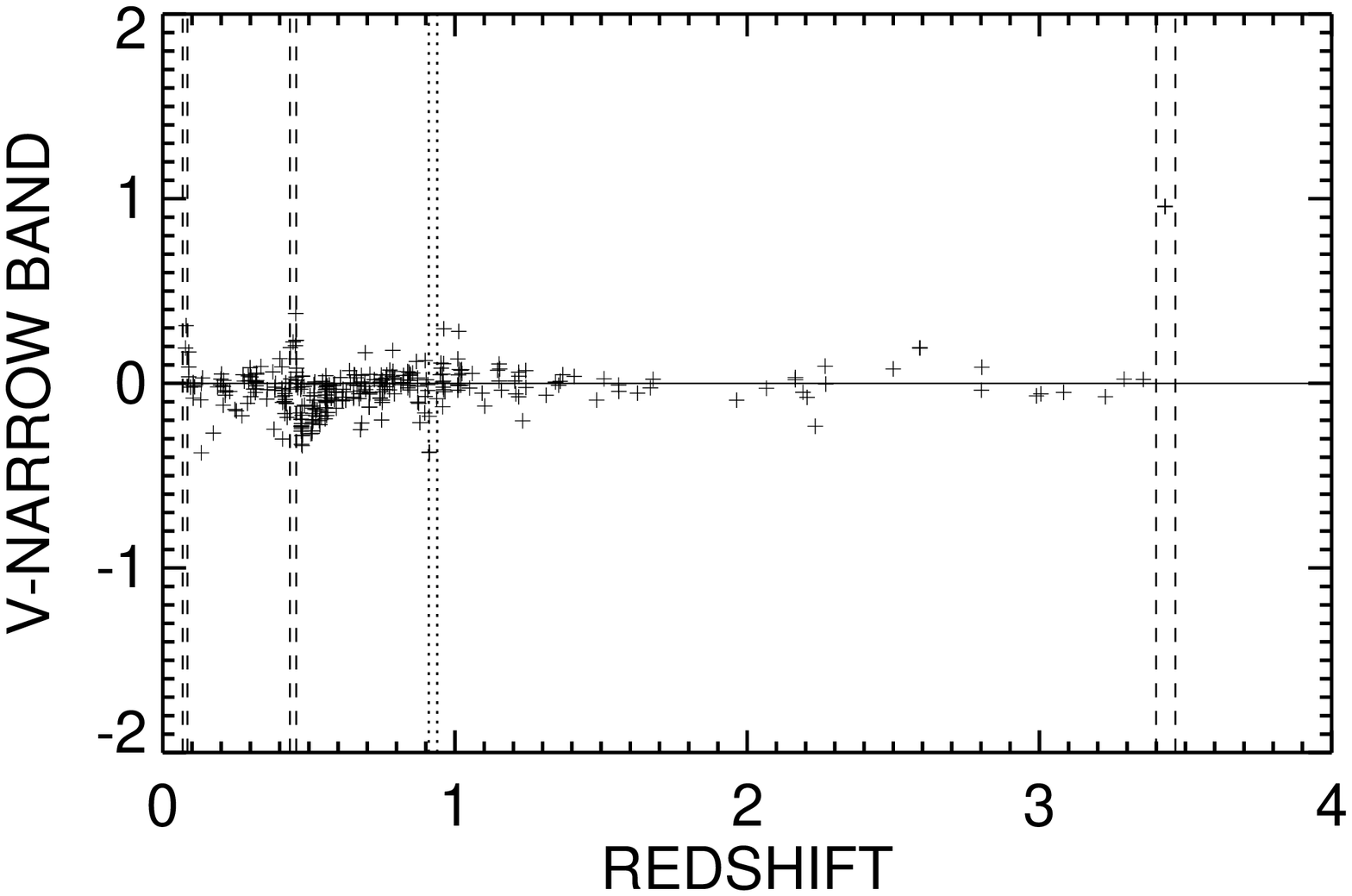}
\caption{Excess emission in narrow-band over $V$-band vs redshift for
objects in the HDF and SSA22 fields.  Objects with spectroscopic
identifications in the literature (Lowenthal \etal\protect{\markcite{low97}}
1997; Cohen \etal\protect{\markcite{cohenhdf}} 1997; 
Songaila\protect{\markcite{hdf_active}} 1997 HDF Active Catalog web page; 
Steidel \etal\protect{\markcite{stei96a}} 1996a), over the HDF and SSA22 
fields sampled by the LRIS deep imaging fields are shown here.  The
positions and redshift ranges corresponding to features such as
\mgii\ ({\it dotted lines}) and \oiii, \oii, and Ly$\alpha$ ({\it dashed
lines}) falling within the filter bandpass are indicated.  The structure of
the continuum near the \oii\ feature may be seen reflected in both the
`dip' and dispersion of points in this region, and it may also be seen that
emission-line objects in \oii\ and a few \oiii\ emitters may be detected.
Underlying continuum features can also increase the dispersion of these
points in different redshift regions.  At higher redshifts a
correspondingly wider redshift region is sampled by the filter bandpass.
Redshifts used only include values known at the time of the imaging
observations, and do not include follow-up spectroscopic confirmations of
candidate Ly$\alpha$ emitters, which will be described in Hu
\etal\protect{\markcite{hu98}} (1998).\label{fig:5}}
\end{figure}

The redshift information also allows us to check the equivalent width
selection procedure.  In Fig.~\ref{fig:5} we show \vmn\ vs redshift, 
with the redshift intervals matching the narrow-band filter band-pass
indicated for emission from \oiii\ $\lambda\,5007$ (at $z\sim 0.08$),
\oii\ $\lambda\,3727$ (at $z\sim 0.45$), \mgii\ $\lambda\,2800$ (at $z\sim 
0.93$), and Ly$\alpha$ $\lambda\,1216$ (at $z\sim 3.43$).  These are the 
most prominant spectral features in most galaxies (cf., Songaila
\etal\markcite{ksurvey_3} 1994; Cowie \etal\markcite{gal_form}
1995a), with \mgii\ normally seen in absorption, and they are clearly
seen in the color-redshift plot.  The increasing width of the
redshift interval corresponding to the narrow-band wavelengths at
high redshifts is also seen.  The scatter of points gives an idea of
the real distribution of the signal strength of each feature, and it
may also be seen that the high-$z$ Ly$\alpha$ emitter, corresponding to the
Lowenthal \etal\ object, does indeed have appreciably higher
equivalent width than any of the \oii\ emitters.  We also note that
other structures arise from additional spectral features (\eg, the
4000 \AA\ break and \feii\ absorption lines) which add to the
dispersion.

\begin{figure}[h]
\epsscale{0.90}
\plotone{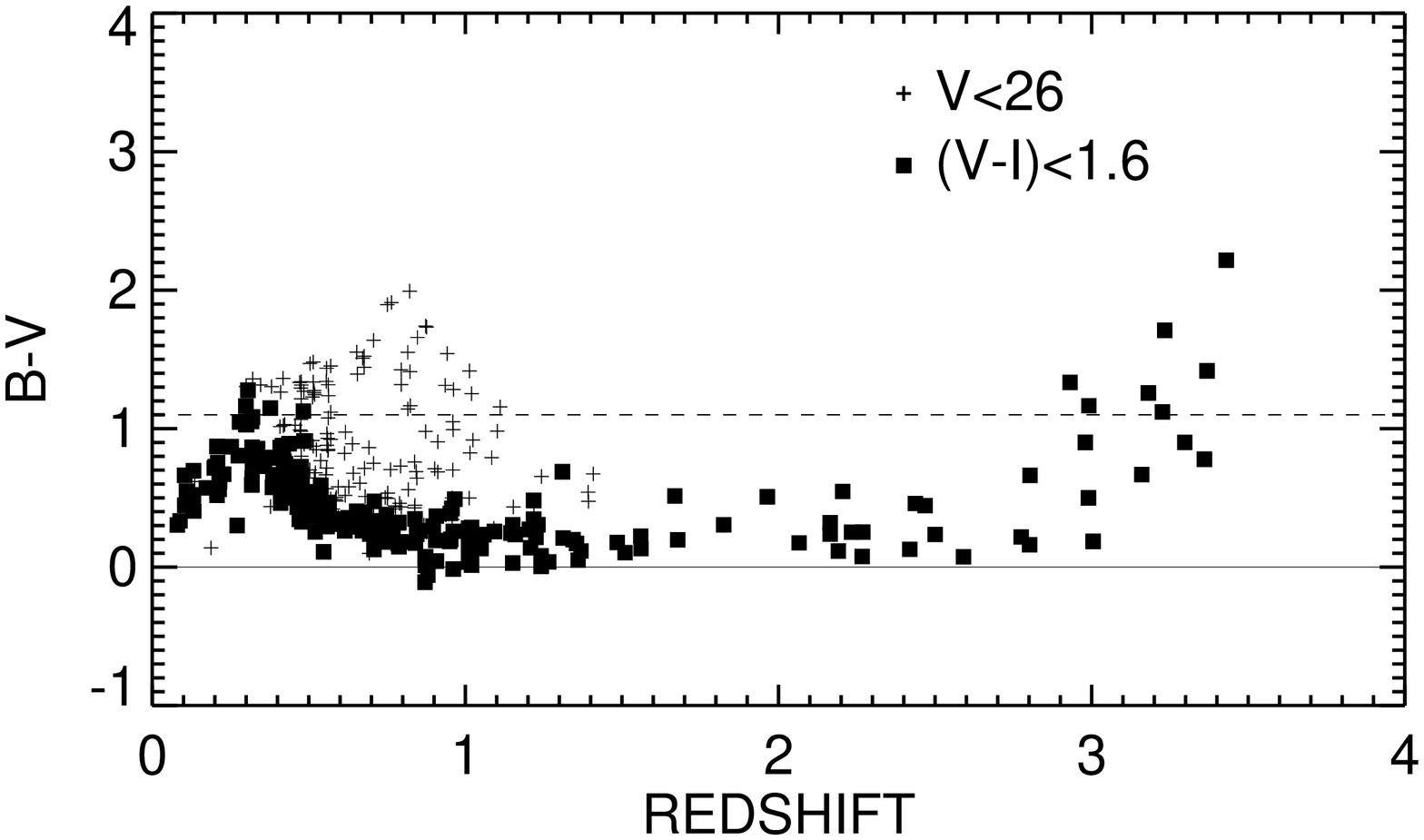}
\caption{\bmv\ vs redshift for SSA22 and the HDF.  For
objects with known redshifts in these two fields, the $V < 26$ sample
({\it pluses}) is divided according to \vmi\ color.  For those objects
which are relatively blue in \vmi\ colors [\vmi\ $< 1.6$] ({\it filled
squares}) a clear trend in \bmv\ color with redshift may be seen,
corresponding to passage of continuum break features through the reference
bandpasses.  At high redshifts ($z\gtrsim2.9$) star-forming galaxies,
which will have flat \vmi\ spectra, appear extremely red in \bmv, and
may be selected by their color breaks.  A dashed reference line at \bmv\
corresponds to our \bmv\ color selection.\label{fig:6}}
\end{figure}

The color trends shown in Fig.~\ref{fig:4} can also be examined in more
detail by supplementing the measured colors for these objects with
redshift information.  In Fig.~\ref{fig:6} we plot \bmv\ colors vs
redshift for all galaxies in our $V$-selected catalog for which this
information is available.  It may be seen that applying a blue \vmi\ color
criterion (\thinspace\vmi\ $ < 1.6$) selects galaxies over a range of
redshifts, extending at least to $z\sim3.5$, but that within this
subsample there are very definite trends in \bmv\ color with redshift.
This function (\thinspace\bmv\ vs redshift) remains double-valued out to
about $z\sim3$, with a progressive decrease (bluing) in \bmv\ color as the
4000 \AA\ Balmer break moves out of the $B$ band to longer wavelengths in
the observed frame. At higher redshifts the combination of the Lyman break
and the depression of the continuum below Ly$\alpha$ due to increasing
absorption by Ly$\alpha$ forest clouds leads to progressively redder observed
\bmv\ colors, and $z>3$ galaxies are easily distinguished.  The data show
that using the color criterion \vmi\ $ < 1.6$ and \bmv\ $> 1.1$ will
select primarily $z \gg 3$ galaxies, though a with a small admixture of
low-redshift galaxies and stars.

\begin{figure}[h]
\epsscale{0.90}
\plotone{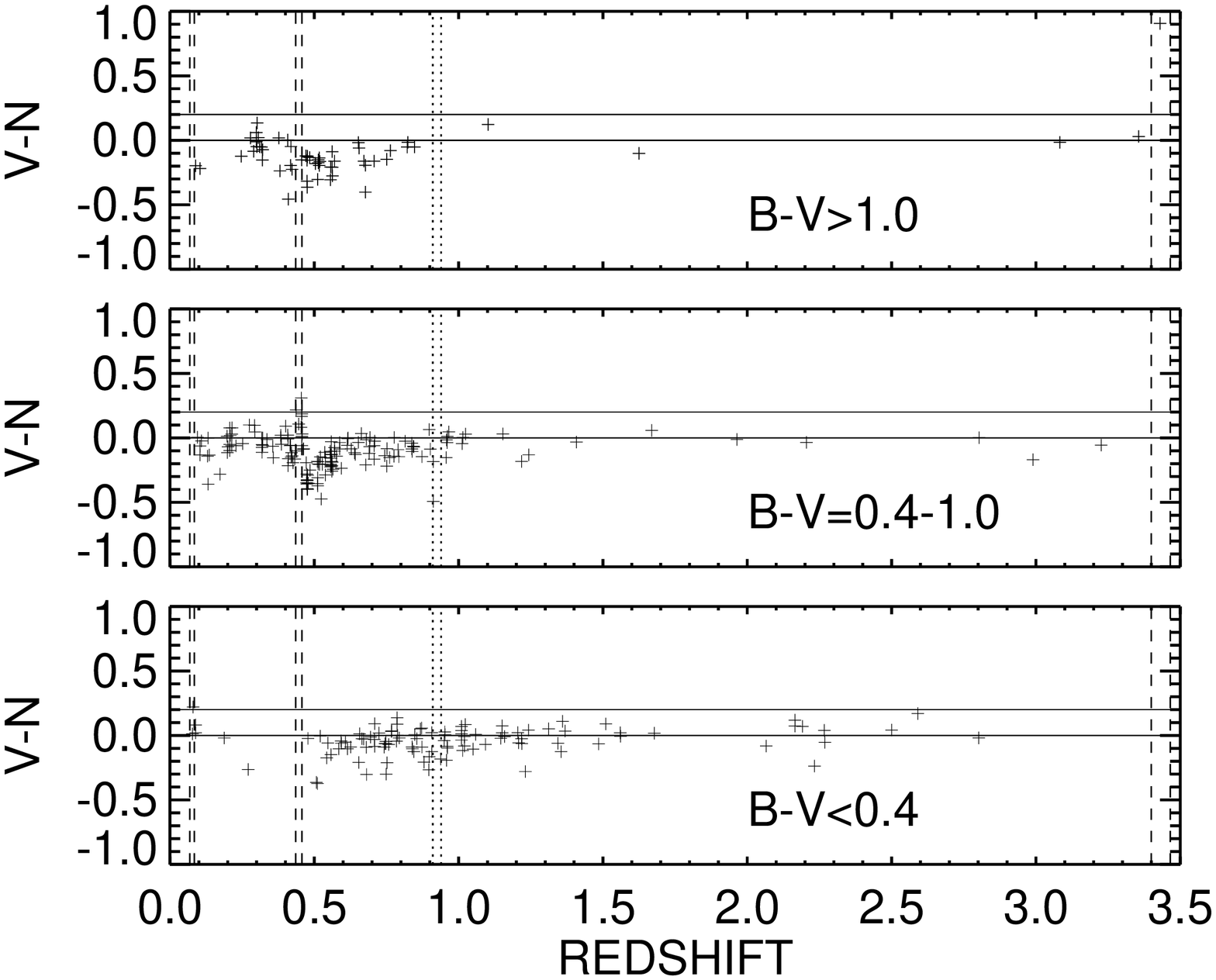}
\caption{\vmn\ vs redshift for objects with $N(AB) < 24.5$ in three
\bmv\ color subsamples.  The data are grouped according to \bmv\ color, and
illustrate the color trends with redshift of Fig.~\protect{\ref{fig:6}}, which
demonstrates that it is possible to discriminate between classes of
emitters using color data, and hence to identify weaker Ly$\alpha$ emitters
than would be possible with a strong equivalent width criterion.  The solid
lines show objects with \vmn\ $> 0.2$.  For \bmv\ $> 1.0$ the identified
emission-line object corresponds to Ly$\alpha$ at $z\sim3.4$.  In the
intermediate $0.4 <$ \bmv\ $< 1.0$ color range \oii\ emission-line objects
are selected. At \bmv\ $< 0.4$ the narrow-band excess objects correspond to
\oiii\ emitters.\label{fig:7}}
\end{figure}

In Fig.~\ref{fig:7} we show how local continuum shape can be used to
discriminate between the three classes of emission lines (\oii,
\oiii, and Ly$\alpha$) that produce significant equivalent widths in the
narrow band and hence to identify weaker Ly$\alpha$ emitters than would
be picked out by the strong equivalent width criterion used
previously.  For the low-redshift \oiii\ lines the continuum is flat
and \bmv\ $< 0.4$ ({\it lower panel}).  For the \oii\ lines there is
a strong discontinuity just redward of 3727 \AA\ (\eg, Lilly
\etal\markcite{lilly_1} 1995) which places \bmv\ in the $0.4-1.0$
range.  Finally, for the Ly$\alpha$ lines the strong discontinuity produced
by the Ly$\alpha$ forest across the Ly$\alpha$ line places \bmv\ $> 1.0$
(Cowie\markcite{cowie88} 1988; Madau \etal\markcite{madau96} 1996).

\begin{figure}[h]
\epsscale{0.90}
\plotone{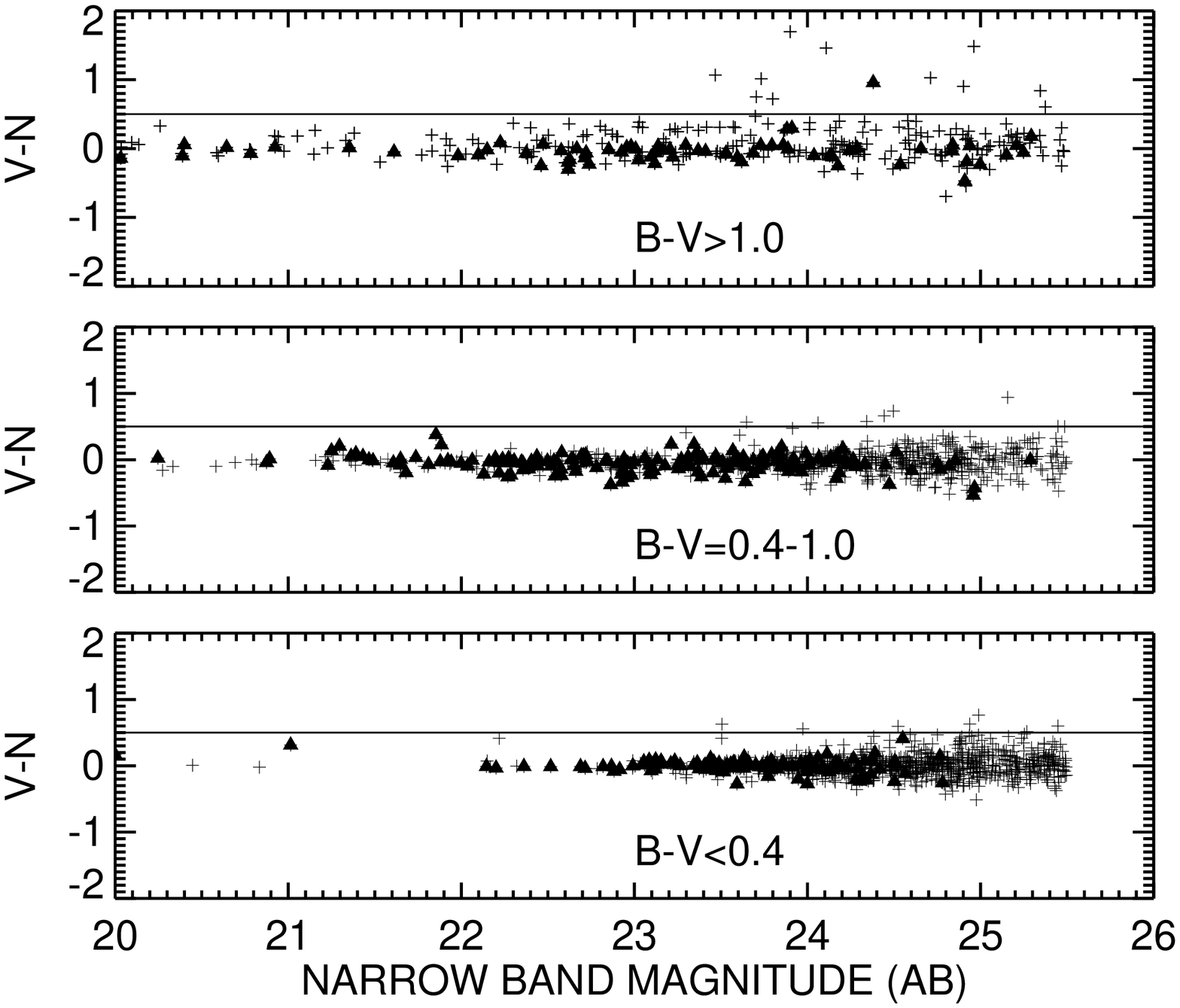}
\caption{\vmn\ vs $N\/$ for $N(AB) < 25$ in the HDF and
$N(AB) < 25.5$ in SSA22 divided by \bmv\ color into three bins.
Objects with known redshifts are marked with triangles.  The dividing
line in \vmn\ is set at 0.5.\label{fig:8}}
\end{figure}

We have divided the entire sample ($N$($AB$)$ < 25$ in the HDF, $N$($AB$)$
< 25.5$ in SSA22) in the same way in Figure~\ref{fig:8}.  Here it can be
seen that the \bmv\ selection chooses out the highest equivalent width
objects with only one object falling into the \bmv\ $< 1$ class having
\vmn\ $> 0.78$.  As is illustrated in Fig.~\ref{fig:8} a \vmn\ $> 0.5$,
\bmv\ $> 1$ criterion should be effective in picking out emitters.
However, weaker \vmn\ cuts become contaminated by error and color spread.
The color-enhanced selection criterion therefore yields only a small number
of additional objects over the simple strong equivalent width criterion.

\section{Discussion}

The strong equivalent width criterion gives ten strong Ly$\alpha$ emitters with
flux $> 2 \times 10^{-17}$ erg cm$^{-2}$ s$^{-1}$ in the 46 $\sq'$
field coverage (about $800/\sq\arcdeg$) lying in the redshift range
$z=3.405 \to 3.470$, or roughly 13,000/unit $z/\sq\arcdeg$.  Two of
these objects --- both lying in the SSA22 region --- are completely
undetected in the continuum at the $1\ \sigma$ level of $V = 27.5$, while
the remainder have $V$ magnitudes ranging from 24.6 to 26.5 (Table
\ref{tbl-1}).  To a $V=25.5$ magnitude limit a color criterion \bmv\ $>
1.1$, \vmi\ $ < 1.6$ of the type discussed above gives 72 objects within
the sample area when known low-redshift galaxies or stars are excluded.
If we assume a rough $z$ range of $3.1 \to 3.5$, where the upper limits is
determined by the passage of the forest through the middle of the $V$
band, this selection also corresponds to a surface density of 13,000/unit
$z/\sq\arcdeg$.  While the number is quite uncertain because of the
small number statistics, the choice of magnitude limit, and the
possibility of redshift clustering, there are comparable numbers of strong
Ly$\alpha$ emitters and color-selected galaxies at these flux selection limits.
This is consistent with the spectroscopic properties of the color-selected
objects with measured high redshifts, and it appears that Ly$\alpha$ emission is
quite common in the high-redshift objects.  In addition, there appear to
be numbers of bare Ly$\alpha$ emitters where the continuum is undetectable,
which balances to some extent the presence of color-selected objects
without strong emission.

In the absence of extinction and with the assumption that the Ly$\alpha$ 
emission is produced
by photoionization by stars, the conversion from Ly$\alpha$ luminosity to
star formation rate is 1 \msun\ yr$^{-1}\ \to 10^{42}$ erg s$^{-1}$ in
the Ly$\alpha$ line, where we have used Kennicutts's\markcite{kenn83} (1983)
relation between the star formation rate and H$\alpha$ luminosity and the
Case B Ly$\alpha$/H$\alpha$ = 8.7 (\markcite{brock}Brocklehurst 1971).  Because 
of the effects of extinction the Ly$\alpha$ luminosity is likely to
underestimate the SFR, while if some portion of the Ly$\alpha$ luminosity is
powered by AGN we will overestimate the SFR.  The size of these effects is not
easy to estimate, though the rest-frame equivalent widths of the lines (Table
\ref{tbl-1}) imply that if the objects are photoionized that extinction cannot
reduce the line luminosities by more than a factor of about two before
reasonable theoretical upper bounds on the equivalent width (Charlot \&
Fall\markcite{charl93} 1993) are exceeded.  We also note that in the spectra
obtained to date of the SSA22 emitters (Hu \etal\markcite{hu98} 1998) there is
little sign of the \civ\ line which might indicate AGN activity.

\begin{figure}[h]
\epsscale{0.85}
\plotone{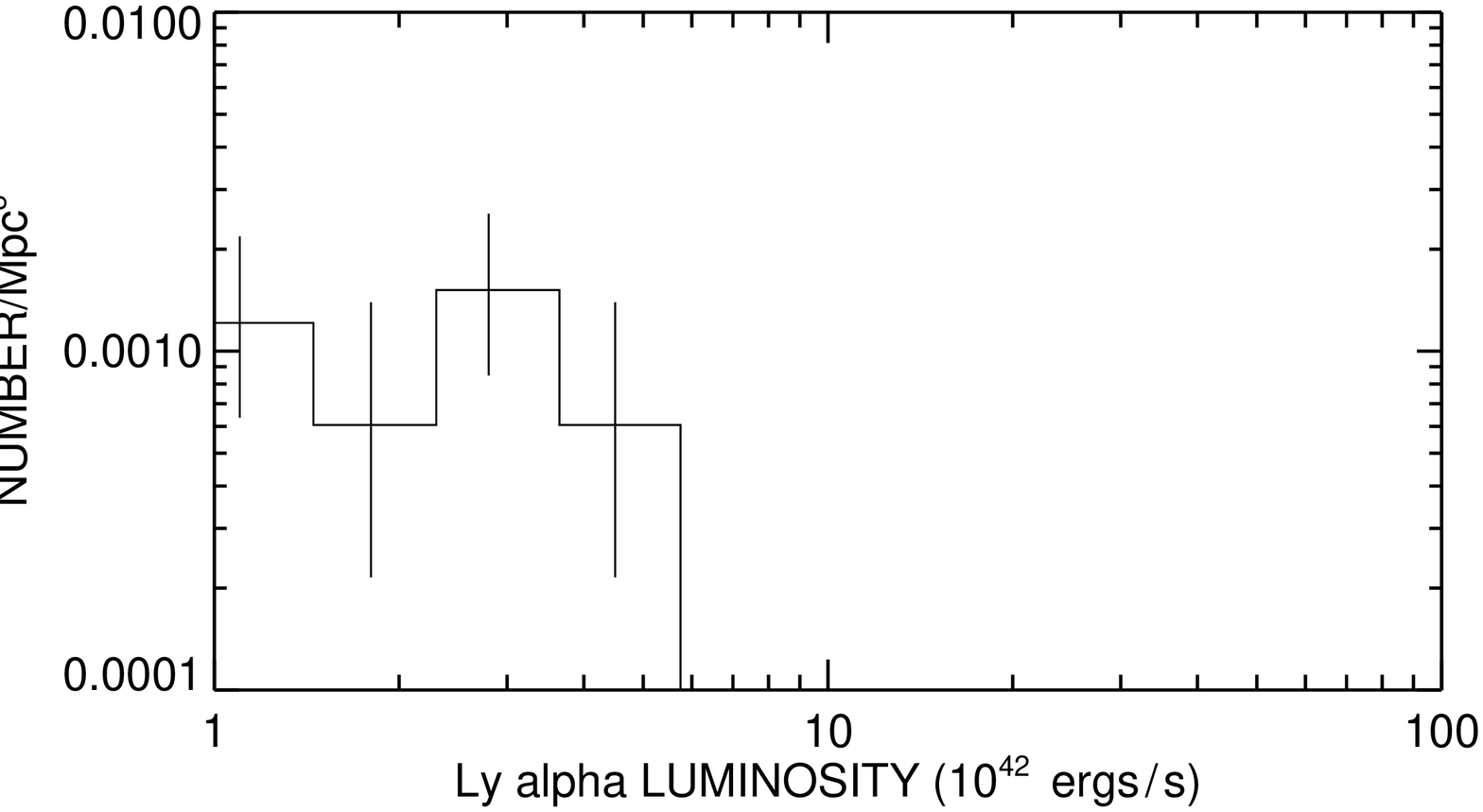}
\caption{Distribution of Ly$\alpha$ emission vs Ly$\alpha$
luminosity for the strong equivalent width emission-line objects
in the SSA22 and HDF LRIS fields.  The horizontal axis may
be converted to star formation rate in the absence of extinction
and assuming that the Ly$\alpha$ is produced by photoionization, whence
$10^{42}$ erg s$^{-1} \approx 1$ \msun\ yr$^{-1}$.  The errors
are $\pm1\ \sigma$ based on the number of objects in each bin.
See text for a more extensive discussion.\label{fig:9}}
\end{figure}

Proceeding therefore under the assumption that the line is photoionized
by stars and neglecting extinction, we find maximum star formation
rates of just under $10\ h_{50}^{-2}$ \msun\ yr$^{-1}$ for
$q_0 = 0.5$, which are very similar to those seen at lower $z$
(\markcite{mdot}Cowie \etal\markcite{mdot} 1997) and those inferred
from the continuum light of color-selected objects at these redshifts
(Steidel \etal\markcite{stei98} 1998).  In Fig.~\ref{fig:9} we show the
Ly$\alpha$ luminosity function constructed for $H_0 = 65\ {\rm km\ s}^{-1}\
{\rm Mpc}^{-1}$ and $q_0 = 0.5$.  In the absence of extinction this may be
directly converted to an $\dot{M}$ distribution using the
$L$(Ly$\alpha$)--$\dot{M}$ relation.  The $\dot{M}$ function is quite similar
to that seen at $z=1$ (Cowie \etal\markcite{mdot} 1997) within the
large statistical uncertainties and modulo the different methodologies
used (UV fluxes vs Ly$\alpha$ luminosities).  It is more probable that
$\dot{M}$ is underestimated at the high redshift because extinction 
must have a larger effect on the Ly$\alpha$ line than on the
continuum.  Integrating through the $\dot{M}$ function we find a star
formation density of 0.01 \msun\ Mpc$^{-3}\ {\rm yr}^{-1}$, which is
quite similar to that inferred from color arguments (\eg, Madau
\etal\markcite{madau96} 1996, Mobasher \& Mazzei\markcite{mob97}
1997).  Again, it must be emphasized that this is an extreme lower
limit since we neglect extinction.

\acknowledgements

We thank Boris Shnapir for assistance in the design and fabrication of
the narrow-band filter, Bob Williams and Richard Ellis for comments
on an earlier draft of this paper, and Tom Bida and Bob Goodrich for their
assistance in obtaining the observations, which would not have been
possible without the LRIS spectrograph of Judy Cohen and Bev Oke.
Support for this work was provided by the State of Hawaii and by NASA
through grants number GO-5975.01-94A, GO-06222.01-95A, and AR-06377.06-94A 
from the Space Telescope Science Institute, which is operated by AURA, Inc.,
under NASA contract NAS 5-26555.  E.M.H.\ would also like to gratefully
acknowledge a University Research Council Seed Money grant.

%\newpage
% Dummy tables for cross-referencing
%\begin{table}
%\tablenum{1}
%\dummytable\label{tbl-1}
%\end{table}

%
% Table 1
%
\begin{deluxetable}{lcccccc}
\tablenum{1}
%\label{tbl-1}
%\tablewidth{42pc}
\tablecaption{Properties of Strong Emitters\label{tbl-1}}
\tablehead{              &
\colhead{Flux}           & \colhead{W$_{\lambda}$(observed)} & 
                         &                                   & 
                         &                                   \\[0.5ex]
\colhead{ID}             &
(erg cm$^{-2}$ s$^{-1}$) & (\AA)                             & 
\colhead{~$V$}           & \colhead{($V - I$)}               & 
\colhead{($B - V$)}      & \colhead{$z$}    
}
\startdata
\sidehead{HDF, $N<25$}\tableline\tablevspace{2pt}
\phn\phn{HDF--LA1} & $8.0\,(-17)$ & 140 & 24.6 & 0.9 & 1.1 & \nodata\nl
\phn\phn{HDF--LA2} & $6.1\,(-17)$ & 135 & 24.8 & 0.5 & 1.4 & \nodata\nl
\phn\phn{HDF--LA3} & $5.2\,(-17)$ & 250 & 25.7 & 0.0 & 2.2 & \nodata\nl
\phn\phn{HDF--LA4} & $3.2\,(-17)$ & 115 & 25.4 & 1.6 & 1.9 
                    & 3.430\tablenotemark{a}\nl
\phn\phn{HDF--LA5} & $2.4\,(-17)$ & 250 & 26.5 & 0.2 & 1.1 & \nodata\nl
\tableline
\noalign{\vskip2pt}
\sidehead{SSA22, $N<25.5$}\tableline\tablevspace{2pt}
\phn\phn{SSA22--LA1} & $6.7\,(-17)$ & 325 & 25.7 & 0.8 & 2.4 
                    & 3.455\tablenotemark{b}\nl
\phn\phn{SSA22--LA2} & $5.2\,(-17)$ & 230 & 25.6 & 0.8 & 1.6  
                    & 3.460\tablenotemark{b}\nl
\phn\phn{SSA22--LA3} & $3.8\,(-17)$ & $>845\,\phm{>}$ 
                    & $-28.0$\tablenotemark{c}\phm{-$^c$} 
                    & \nodata & \nodata & 3.450\tablenotemark{b}\nl
\phn\phn{SSA22--LA4} & $2.5\,(-17)$ & 140 & 25.9 & 0.2 & 1.3 & \nodata\nl
\phn\phn{SSA22--LA5} & $2.0\,(-17)$ & $>405\,\phm{>}$ 
                    & $-28.7$\tablenotemark{c}\phm{-$^c$}
                    & \nodata & \nodata & \nodata\nl
\phn\phn{SSA22--LA6} & $1.5\,(-17)$ & 115 & 26.2 & 0.4 & 1.0 & \nodata\nl
\phn\phn{SSA22--LA7} & $1.2\,(-17)$ & \phn90 & 26.2 & 2.2 & 1.2 
                    & 0.442\tablenotemark{b}\nl
\enddata
\tablenotetext{a}{Lowenthal \etal\ 1997}
\tablenotetext{b}{Hu \etal\ 1998, in preparation}
\tablenotetext{c}{A negative $V$ magnitude here indicates that there is a 
negative flux in the aperture. The $1\ \sigma$ limits on $V$ are
27.4 (HDF) and 27.5 (SSA22).}
\end{deluxetable}

\begin{figure}[h]
\figurenum{1a}
\caption{Narrow-band image of the HDF and surrounding
field. The displayed field of view shows a $380'' \times 275''$ region
from a 2-hr exposure taken with LRIS on the Keck\thinspace II 10-m
telescope through the 5390/77 \AA\ narrow-band filter in the collimated
beam.  The composite image quality is $\sim 0\farcs7$ FWHM, with a
$1\ \sigma$ limiting $AB\/$ magnitude of 26.8 over a $3''$ diameter
aperture, corrected to a total magnitude following Cowie 
\etal\protect{\markcite{cowie_1}} (1994), and with a
corresponding $1\ \sigma$ flux of $6 \times 10^{-18}$ erg cm$^{-2}$
s$^{-1}$ over the same aperture.  Circled items indicate field objects
with narrow-band equivalent widths in excess of 77 \AA\ in the observed
frame, and which also fall within the central $6\farcm0 \times 4\farcm2$
region used for the narrow-band catalog (including objects brighter
than the $5\ \sigma$ limiting magnitude of 25).  The heavily circled
object is the one catalogued narrow-band object which lies within the
HDF proper.  (The LRIS field displayed covers an area 4.4 times larger
than the HDF field).\ \ This has a redshift identification by
Lowenthal \etal\protect{\markcite{low97}} (1997) of $z=3.430$ (object
id hd2-0698-1297), in agreement with the Ly$\alpha$ selection of the
filter.  This is also the only object among currently reported galaxies
lying within the HDF with a redshift that would place Ly$\alpha$
emission within the filter bandpass; however, we note that no
information about object emission-line redshifts in the selected
bandpass was available at the time the filter was designed. Circled
overlays correspond to a $4\farcs6$ diameter aperture.  Most of the
indicated emission-line objects are more compact than the object
identified by Lowenthal \etal\protect{\markcite{low97}}, as was
generally true of the other high-redshift objects discussed in that
paper.  This object also lies at the lower end of our equivalent width
selection criterion.\label{fig:1a}}
\end{figure}
\begin{figure}[h]
\figurenum{1b}
\caption{Narrow-band image of the SSA22 field.  The
field of view seen in this image is $390'' \times 280''$, and reaches a
$1\ \sigma$ flux of $3\times 10^{-18}$ erg cm$^{-2}$ s$^{-1}$ over a
$3''$ diameter aperture for the 5-hr exposure.  The 
circled items correspond to objects whose emission is in excess of an 
equivalent width of 77 \AA\ for $N(AB) < 25.5$.\label{fig:1b}}
\end{figure} 
\begin{figure}[h]
\figurenum{2a}
\caption{The corresponding $V$-band image of the HDF, comprising 0.6 hrs on
Keck\thinspace II (LRIS) combined with 11 hrs on CFHT with the UH8K mosaic
CCD camera.  The $1\ \sigma$ (Johnson) $V\/$ limit over a corrected $3''$
diameter aperture is 27.4.  Encircled objects are selected from $V < 25.5$
objects which have red \bmv\ colors (\thinspace\bmv$ \geq 1.1$) and blue
\vmi\ colors (\thinspace\vmi\ $\leq 1.6$), many of which should correspond to
high-redshift galaxies.  Again, objects with heavy circles indicate galaxies
which meet the selection criteria, and also have measured redshifts,
given in Fig.~\protect{\ref{fig:2b}}.  Four of the remaining color-selected 
objects lie in the HDF proper.  Of these, one (hd2\_0853\_0319) has a tentative 
redshift of $z=3.35$ from Lowenthal \etal\protect{\markcite{low97}} 1997.
\label{fig:2a}}
\end{figure}
\begin{figure}[h]
\figurenum{2c}
\caption{
The $V$-band image of SSA22, with circled objects showing the color
selection for high-redshift galaxies applied in Fig.~\protect{\ref{fig:2a}},
and with those galaxies with measured redshifts (known in the literature
prior to our spectroscopic follow-up) indicated with heavy
circles.  The $1\ \sigma$ $V\/$ limit is 27.5.\label{fig:2c}}
\end{figure}
\newpage
\begin{figure}
\vbox to 2.9in{\rule{0pt}{2.9in}}
\includegraphics{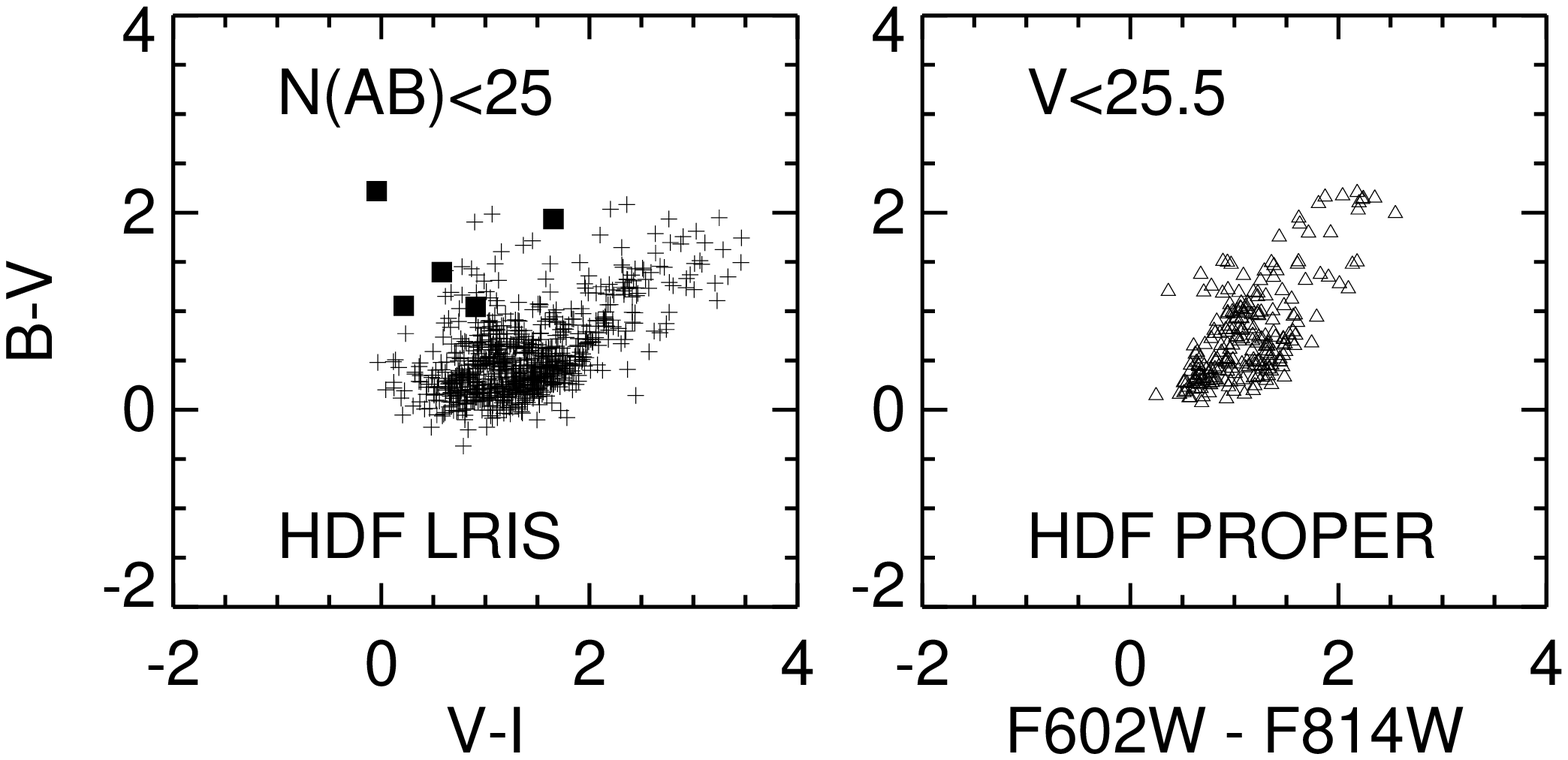}
%\refstepcounter{figure}               % turn off Fig. label
\label{fig:anonymous}
\end{figure}

\end{document}